\documentclass[epj]{svjour}

\usepackage{amsmath,amsfonts,amssymb,graphicx,url}
\usepackage{fancyhdr}
\def\makeheadbox{{
 }}

\def\mytitle{Statistics in the Landscape of Intersecting Brane Models}
\def\myauthors{Florian Gmeiner}
\def\mytype{Contributed Talk}    
\def\mysession{Theoretical Models}

\begin{document}
\title{Statistics in the Landscape of Intersecting Brane Models}
\author{Florian Gmeiner\thanks{\emph{Email:} fgmeiner@nikhef.nl}}
\institute{NIKHEF, Kruislaan 409, 1098 SJ Amsterdam, The Netherlands}
\date{October 11, 2007}
\abstract{
\vskip -14em\rightline{NIKHEF/2007-023}\vskip 13em
An approach towards a statistical survey of four dimensional supersymmetric
vacua in the string theory landscape is described and illustrated with three
examples of ensembles of intersecting D--brane models.
The question whether it is conceivable to make predictions based on statistical
distributions is discussed.
Especially interesting in this context are possible correlations between low
energy observables.
As an example we look at correlations between properties of the gauge sector of
intersecting D--brane models and Gepner model constructions.
\vskip 1em
{\it Based on a talk presented at ``The 15th International Conference on
Supersymmetry and the Unification of Fundamental Interactions'', July 26 --
August 1, 2007, Karlsruhe, Germany.}
}
\maketitle

\section{Introduction}\label{sec:intro}
As has been known for quite a while but only more widely discussed
recently~\cite{ci:landscape},
string theory provides us with an extremely large number of effective four
dimensional theories.
The main reason for this lies in the abundance of different ways how to
compactify the theory from ten to four dimensions.
This procedure is by no means unique and produces massless
moduli fields in the effective theory.
One way to introduce a potential for these moduli and thereby fix their values
is the use of background fluxes (for a recent review in the context of
intersecting brane models see~\cite{ci:review}), but even after using this
method we are left with an abundance of possibilities.

Facing such a huge number of possible low energy theories, we have to answer
the question why exactly the (supersymmetric) standard model is realised in
nature.
One certainly has to try to identify a selection mechanism within string theory
that singles out a particular solution (e.g. based on
entropy) or one has to face the possibility of retreating to anthropic
reasoning.

However, it might be possible to extract some useful information from a
statistical analysis of solutions by searching for common properties within
ensembles of models at different points in the landscape.
This might give valuable hints for model building by excluding or highlighting
regions of the parameter space.

Moreover, looking for correlations between low energy observables might be an
interesting possibility~\cite{ci:z2,ci:watitalk}.
If found, these correlations could not only teach us some lessons about the
general behaviour of string theory models and help to identify underlying
principles, but they might even lead to concrete results.
If it should turn out that certain correlations exist in a wide variety of
models, one could conjecture them to be true in general and thereby obtaining
testable predictions for experiment.

Besides these interesting, desired correlations, we have to take a different
type of unwanted correlations into account as well~\cite{ci:dilecorr}.
In the analysis of ensembles of models it can be desirable or even
necessary to infer from the properties of a particular subset of models to
the distribution of these features in the whole class.
In the simplest case these subsets are randomly chosen, but most of the time
one has no other choice then to introduce a bias on the basis of which models
are feasible to calculate.
In this case one has to be very careful not to run into unwanted correlations.
Concretely, the expectation value of the statistical distribution might
explicitly depend on the choice function.

How should one proceed to obtain statistical data on string compactifications?
There are basically two routes one could follow. One of them relies on a true
statistical approach, using an approximated measure for the space of
models~\cite{ci:stat}.
The other one uses ensembles of explicitly calculated models at specific points
in the landscape to compute frequency distributions of their properties, which
eventually can be extrapolated to a wider class of compactifications.
This is the approach followed here, in particular we will consider frequency
distributions of properties of intersecting brane
models~\cite{ci:z2,ci:ibms,ci:dt,ci:z6,ci:z6p}.
Other corners of the landscape that have been studied using a similar method
include Gepner models~\cite{ci:gepner,ci:timth}, which we will use later to
compare results on correlations, and orbifold compactifications of heterotic
string theory~\cite{ci:het}.

\section{Intersecting brane models}
\label{sec:ibms}
We will work with simple orientifold models of type IIA, compactified on
toroidal orbifold backgrounds and equipped with D6--branes at
angles~\cite{ci:bdl}.
These models have been studied in great detail over the last years and are
still being used for phenomenological model building~\cite{ci:ibmrev}.
However, to be able to study large quantities of different constructions (or
even all possible models in a given background), some important points will not
be taken into account. One of them is the question of moduli stabilisation, in
particular the inclusion of background fluxes, another one are contributions of
non--perturbative effects, such as instanton corrections.
We believe however, that the frequency distributions obtained in this
simplified setup can be used as a basis for a refined analysis and most of
their properties are not going to change qualitatively.

Specifically, the backgrounds we consider are of the form
\begin{equation}\nonumber
\mathbb{R}^{1,3}\times M,\quad M=T^6/G,\quad
G\in\{\mathbb{Z}_2\!\times\!\mathbb{Z}_2,\mathbb{Z}_6,\mathbb{Z}'_6\},
\end{equation}
where $\mathbb{Z}_6$ and $\mathbb{Z}'_6$ denote two different embeddings
of the group action into the torus--lattice.
The orientifold projection consists of dividing out worldsheet parity,
accompanied by a reflection along three of the torus axes in space--time.
This introduces topological defects, which are described by orientifold
O6--planes that carry tension and are charged under the RR--seven--form.

To account for the tadpoles introduced by the O6--planes, one introduces
D6--branes in the background, which are space--time filling and wrap,
as the orientifold planes, Lagrangian three--cycles in the compact space.
These cycles can be parametrised using a basis of $H_3(M,\mathbb{Z})$.
It can be split into one part that comes from those torus cycles that survive
the projections (called \emph{bulk cycles} in the following) and another part
that is due to the existence of \emph{exceptional cycles} at the fixed
points of the orbifold action, which are related to the possible blow--up
modes.

For the three geometries under consideration the situation is rather different.
In the case of $T^6/(\mathbb{Z}_2\!\times\!\mathbb{Z}_2)$, we do not have
exceptional cycles, such that the basis consists of torus cycles only. In the
other two cases exceptional cycles do exist and combine with the torus cycles
to fractional cycles.
The existence of exceptional cycles makes a huge difference for the statistics,
as we will see below.

Not every possible background will lead to a valid model. In order to obtain
consistent compactifications, there are several consistency conditions that
have to be fulfilled. Two of them have already been alluded to, namely the
cancellation of tadpoles coming from the RR--seven--form, which reads in
homology
\begin{equation}
\label{eq:tad}
\sum_aN_a\left(\pi_a+\pi'_a\right)=L\pi_{O6},
\end{equation}
where $\pi$ are the three--cycles wrapped by the brane $a$, it's orientifold
mirror $a'$ and the O6--planes. $L$ denotes the orientifold charge and
the sum runs over all stacks of D6--branes in the model, each consisting of
$N_a$ branes.
For a background with third Betti number $b_3$ there are $b_3/2$ such
conditions.
Moreover, since we are looking for supersymmetric models in four dimensions,
which leads to the constraint that all three--cycles have to be special
Lagrangian, we have to impose the condition that the calibration form
$\Omega_3$ vanishes when restricted to the three-cycle.
Additionally anti--branes should be excluded from the spectrum,
leading to the condition that the real part of $\Omega_3$ has to be positive,
\begin{equation}
\label{eq:susy}
\left.\mathrm{Im}\Omega_3\right|_{\pi}=0,\quad\mathrm{Re}\Omega_3>0.
\end{equation}
One last constraint comes from K--theory and can be formulated in our setup
as a condition on intersections between all brane stacks and some orientifold
invariant probe--branes,
\begin{equation}
\label{eq:k}
\sum_aN_a\pi_a\circ\pi_{\mathrm{probe}}=0\mod 2.
\end{equation}

More details on the different geometries, consistency conditions and brane
embeddings can be found in the respective papers~\cite{ci:z2,ci:z6,ci:z6p}.

Each stack of $N$ branes carries a gauge group $G(N)$ on it's worldvolume,
where $G\in\{U,Sp,SO\}$, depending on whether the three--cycle it wraps
coincides with the orientifold plane.
Matter arises at the intersection of brane stacks and their orientifold
mirrors. One can distinguish between chiral and non--chiral multiplets and
compute their multiplicities in terms of the intersection numbers between the
relevant cycles.
For two stacks $a$ and $b$ with $N_a$ and $N_b$ branes we
obtain chiral matter in the bifundamental representation
$(\mathbf{N_a},\mathbf{\overline{N}_b})$ with multiplicity $I_{ab}=\pi_a\circ\pi_b$
and in the representation $(\mathbf{N_a},\mathbf{N_b})$ with multiplicity
$I_{ab'}=\pi_a\circ\pi'_b$. In a similar manner one can compute the non--chiral
multiplets~\cite{ci:z6}.
Moreover, each brane might contribute matter in the adjoint, symmetric and
antisymmetric representations of the gauge group $G(N_a)$.

\section{Statistical distributions}
\label{sec:dist}
\begin{figure*}
\includegraphics[width=0.33\textwidth]{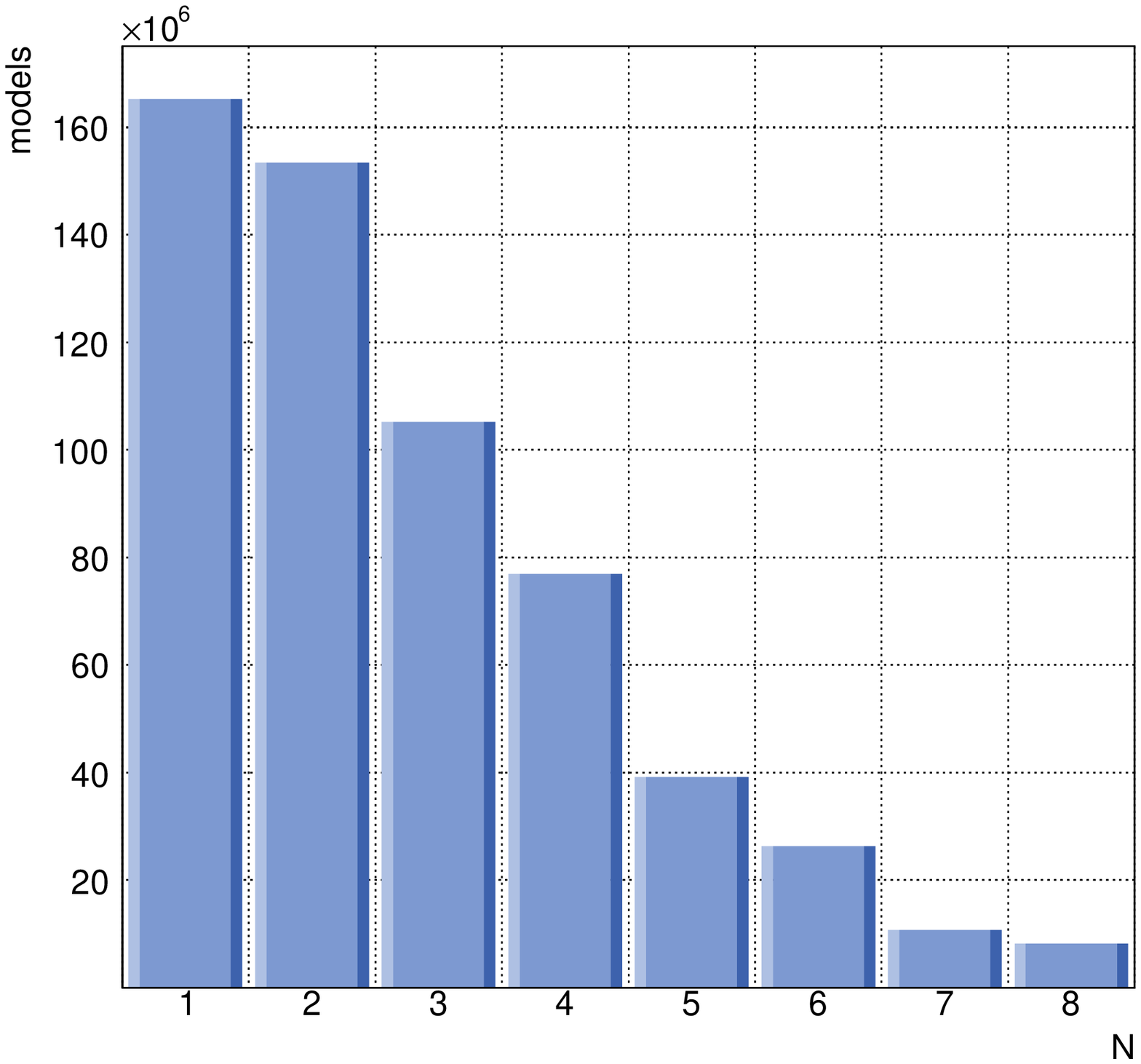}\hfill
\includegraphics[width=0.33\textwidth]{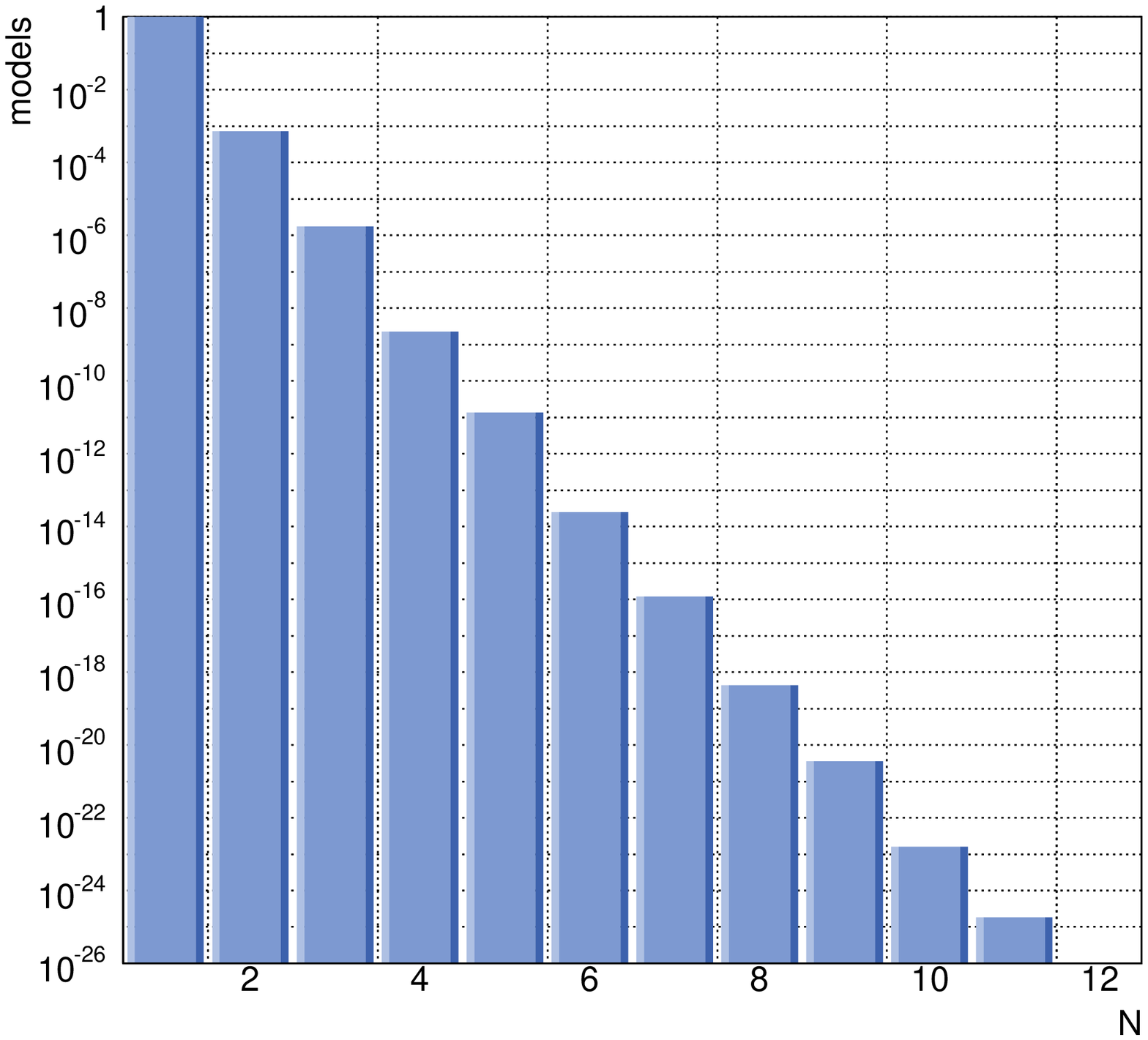}\hfill
\includegraphics[width=0.33\textwidth]{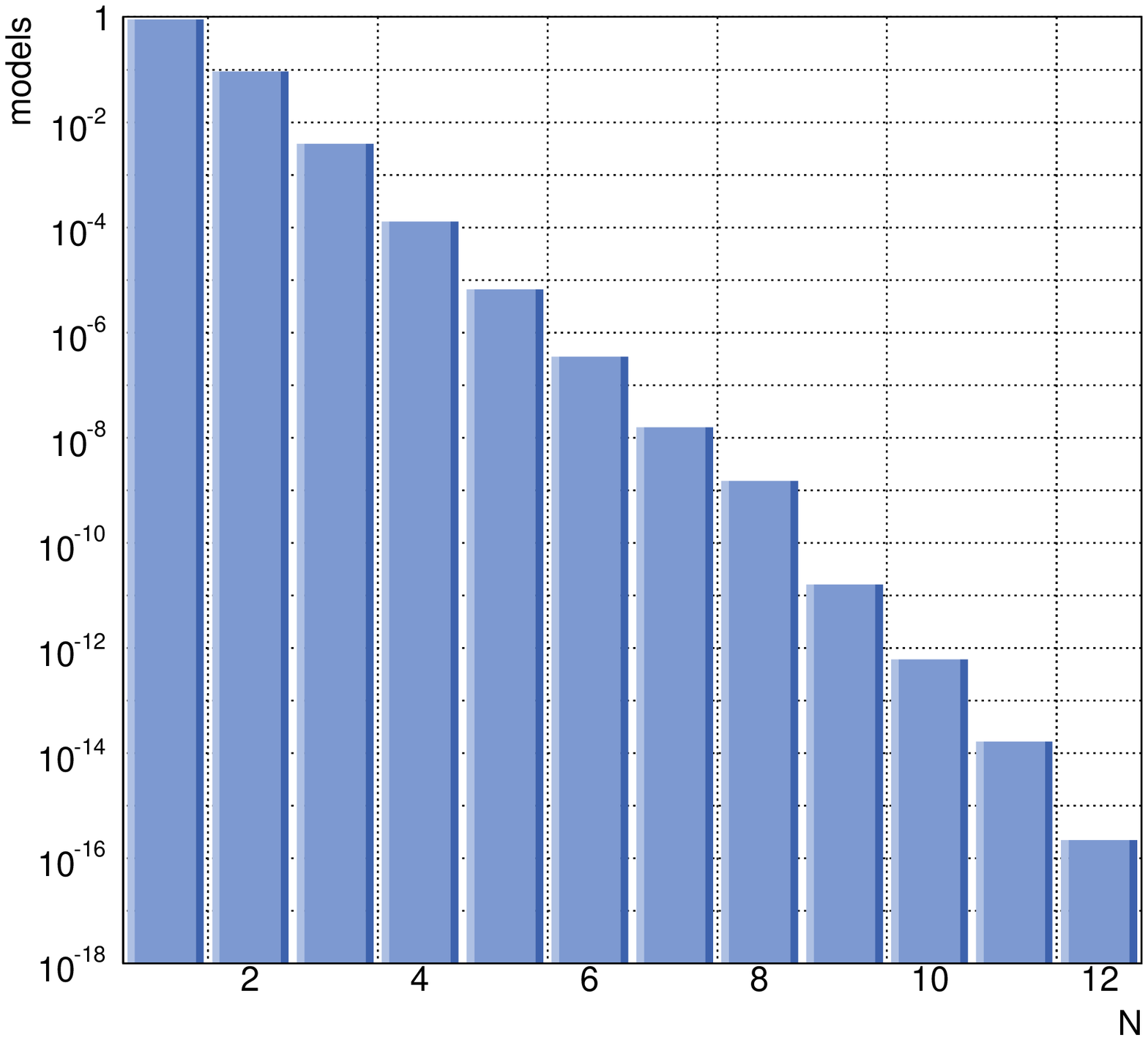}
\caption{Distributions of the probability to find a single gauge group factor
of rank $N$ in the full ensemble of intersecting brane models on
$T^6/(\mathbb{Z}_2\!\times\!\mathbb{Z}_2)$ (left),
$T^6/\mathbb{Z}_6$ (middle) and
$T^6/\mathbb{Z}'_6$ (right).}
\label{fig:un}
\end{figure*}
To evaluate the models described in the last section statistically, large
ensembles of solutions to the constraints~\eqref{eq:tad}, \eqref{eq:susy}
and~\eqref{eq:k} can be
generated using computer algorithms, thereby making use of the fact that after
the introduction of a suitable basis one can express everything in terms of
integer valued algebraic equations.

A complication that arises at this point is that the problem of classifying all
possible solutions to the system of equations is NP--complete~\cite{ci:np}.
However, the number of solutions can be shown to be finite, so it depends on
particular problem whether it is possible to find all possible solutions within
reasonable timescales. In the case of $\mathbb{Z}'_6$ this is indeed possible,
because the set of bulk cycles is very restricted. For the other two
backgrounds under consideration an explicit construction of the full space of
solutions cannot be achieved, which makes it necessary to impose a restriction
to a subset of models. From the properties of the subsets one can
then deduce the frequency distributions for the full set of solutions.

The subsets have to be chosen in such a way that no unwanted bias is introduced
that would distort the statistical distributions.
In the case of $T^6/(\mathbb{Z}_2\!\times\!\mathbb{Z}_2)$ we
have used a cut--off on the space of the real part of the complex structure
which is one of the free parameters. In this way some
interesting models are not captured by the analysis, but
the resulting set of solutions is large enough to obtain valid
statistical distributions\footnote{If one restricts the survey to models with
certain properties, e.g. specific gauge groups, or fixes the number of brane
stacks, statements about the full set of solutions are possible~\cite{ci:dt}.}.
For the $T^6/\mathbb{Z}_6$ background a different method has been used, namely
a restriction to several randomly generated subsets of models, that have been
tested afterwards to make sure that they do not suffer from unwanted
correlations.

The total number of solutions is quite different for the three cases at hand.
For $T^6/(\mathbb{Z}_2\!\times\!\mathbb{Z}_2)$ there are $\mathcal{O}(10^{10})$
models and in the case of $T^6/\mathbb{Z}_6$ and $T^6/\mathbb{Z}'_6$ we find
$\mathcal{O}(10^{28})$ and $\mathcal{O}(10^{23})$ solutions, respectively.
The huge differences, in particular between the first and the latter two
backgrounds, is due to the effect of exceptional branes.
One can show that the tadpole constraints~\eqref{eq:tad} split into a bulk part
and an exceptional part, such that one can treat the two sets of cycles
independently. Each bulk brane can be combined with one of $n_e=2^7$ different
possible exceptional branes to form a fractional cycle, but not all of them
fulfil the consistency conditions, such that the number of possibilities $n_e$
is reduced. For a model with $k$ stacks this amounts to $n_e^k$ combinations.
Not all of these lead to consistent models, but the restrictions from the
exceptional part of the tadpole equations are only polynomial, leading to an
exponential enhancement of the space of solutions.

As an example for the frequency distribution of gauge sector properties, we
will take the probability to find a semi--simple gauge factor of rank $N$
within one model. Combining the probability of several gauge factors, one
can estimate the frequency of certain gauge groups, such as the one of the
standard model, for example.
As one can see clearly from the distributions for the three geometries
(Fig.~\ref{fig:un}), the $T^6/(\mathbb{Z}_2\!\times\!\mathbb{Z}_2)$ orbifold
differs quite dramatically from the other two geometries.
This is again an effect of the contribution of exceptional cycles, which can
be quantified in this case.

For $T^6/(\mathbb{Z}_2\!\times\!\mathbb{Z}_2)$ the distribution for an ensemble
of models with given number of stacks is proportional to
$L^4/N^2$~\cite{ci:dt}.
Including a sum over all possible stack sizes $k$ and, in the case of models
with exceptional stacks, the exponential enhancement factor $n(k)$,
all three distributions can be approximated by
\begin{equation}
\label{eq:unapprox}
P(N) \approx \sum_{k=1}^{L+1-N}\frac{L^4}{N^2}n_e^k,
\end{equation}
where $n_e\equiv 1$ for models without exceptional cycles.
In the case of $T^6/(\mathbb{Z}_2\!\times\!\mathbb{Z}_2)$ we obtain
$(L+1)L^4/N^2-L^4/N$, while for the two embeddings of $\mathbb{Z}_6$
we find $L^4n_e^{(T+1-N)}/N^2$.

\section{Correlations}
\label{sec:corr}
\begin{figure*}
\includegraphics[width=0.33\textwidth]{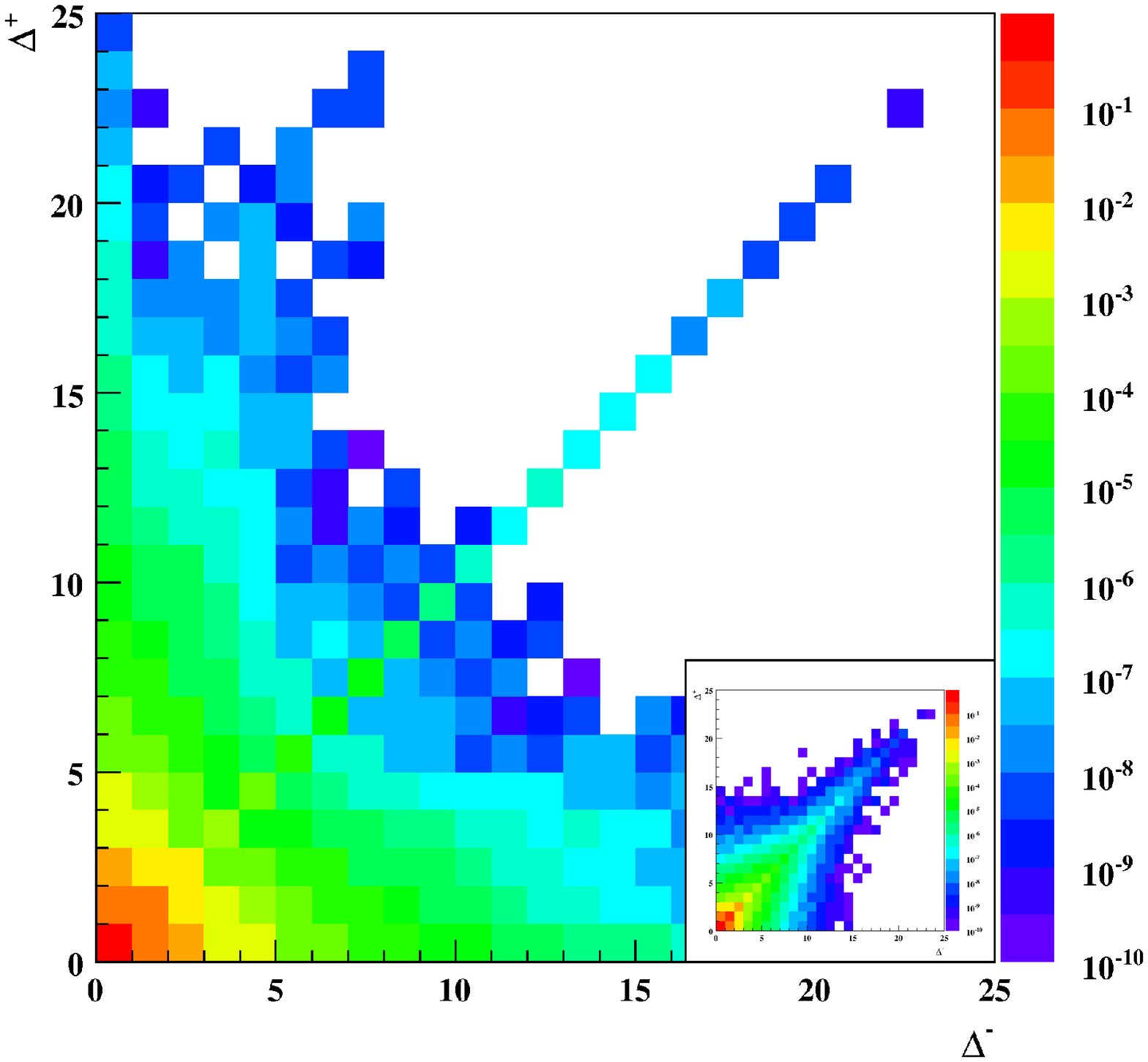}\hfill
\includegraphics[width=0.33\textwidth]{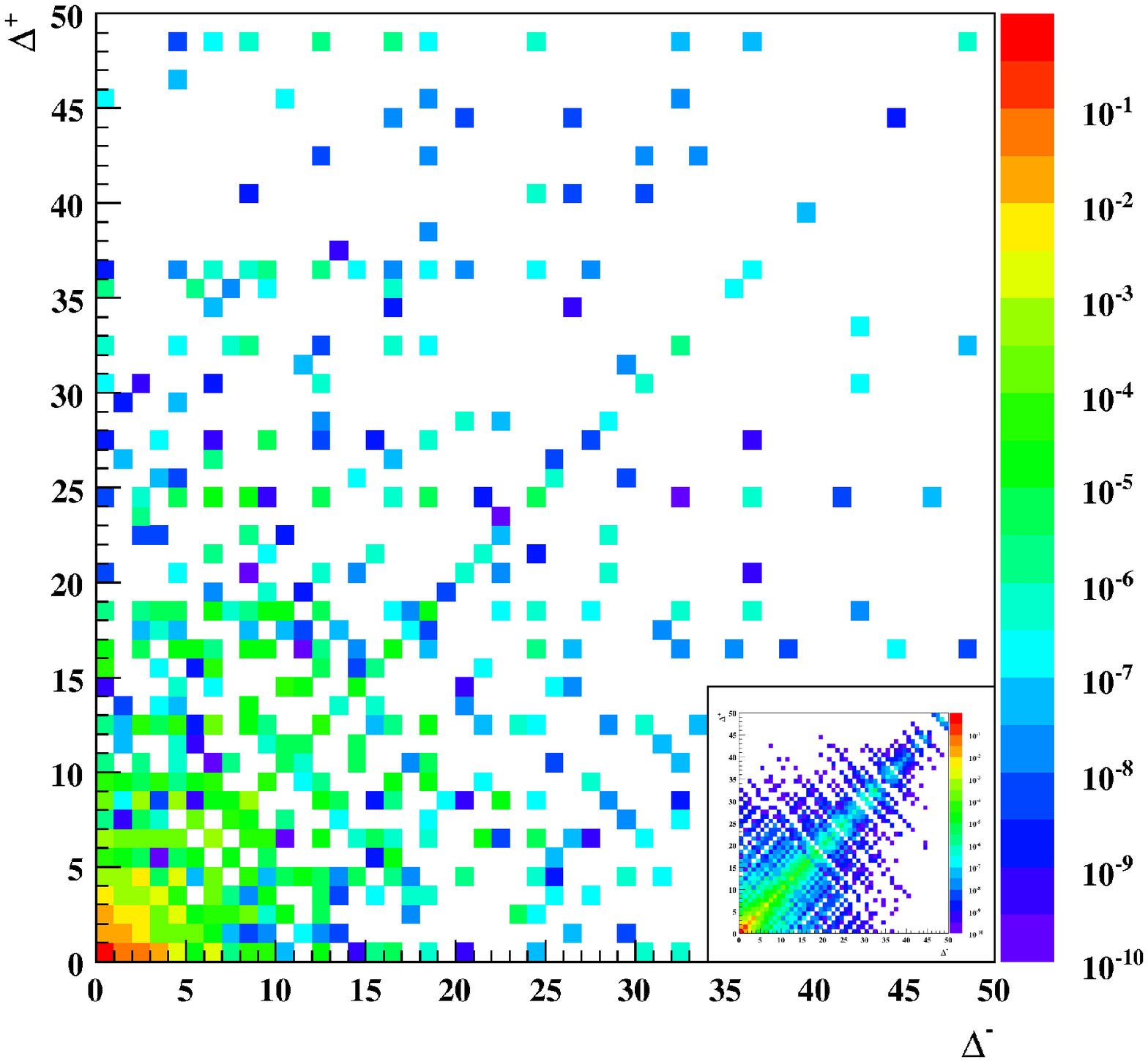}\hfill
\includegraphics[width=0.33\textwidth]{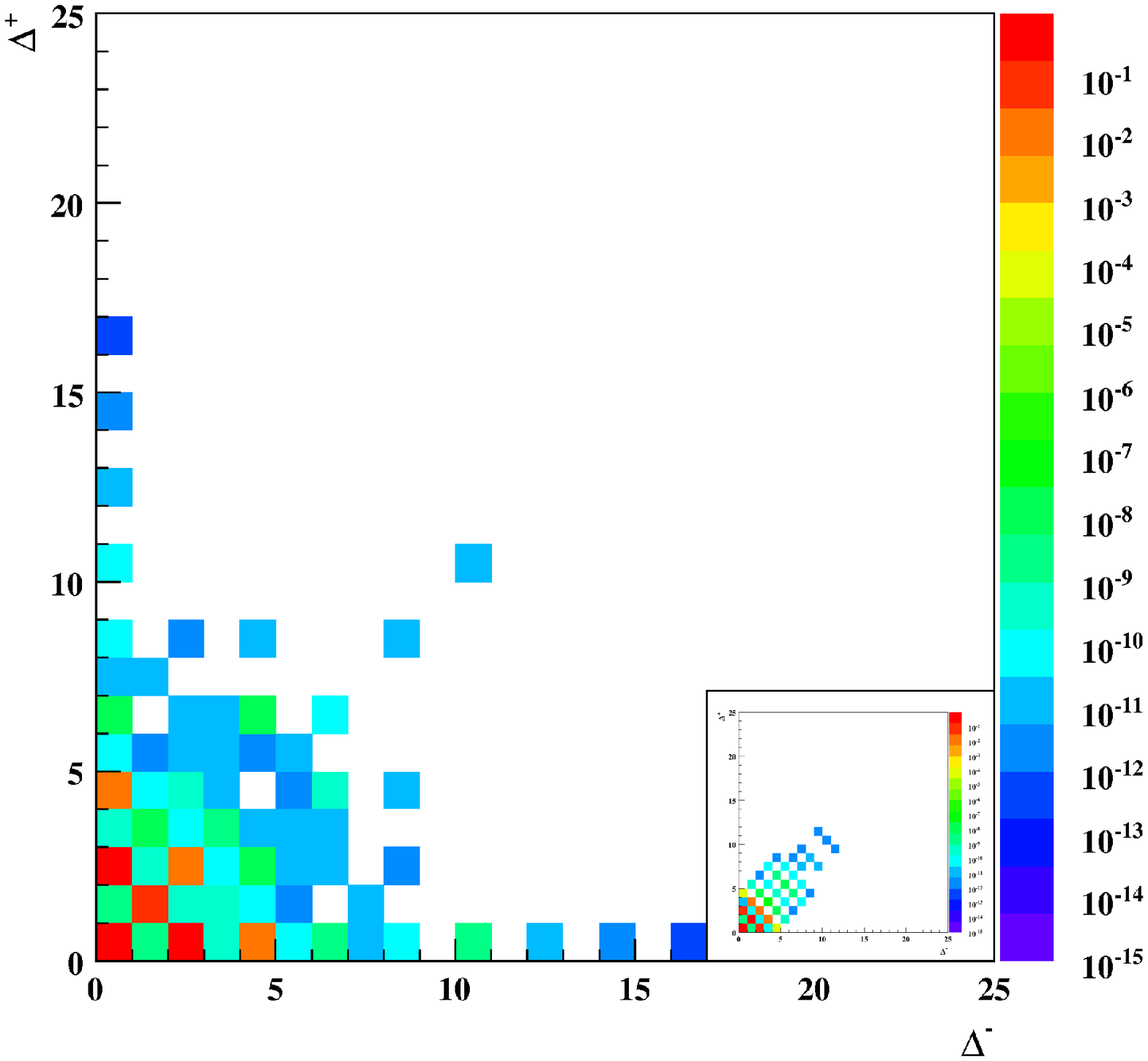}
\caption{Normalised frequency distribution of intersection numbers
$\Delta^\pm$, as defined by~\protect\eqref{eq:deltadef}, for Gepner models
(left) and intersecting brane models on
$T^6/(\mathbb{Z}_2\!\times\!\mathbb{Z}_2)$ (middle) and $T^6/\mathbb{Z}_6$
(right).
The small inserts show the pattern that emerges from randomly generated
intersections using the same set of branes.}
\label{fig:corr}
\end{figure*}
As mentioned in the introduction, the most promising route to obtain
results that may give rise to testable predictions is to look for correlations
between low energy observables.

In the following we will consider only one simple example of correlations
between properties of the gauge sector to show how this might be done in
practise, but certainly many more possibilities could be
considered\footnote{This section contains some preliminary results of work
in progress with Tim Dijk\-stra.}.

Within the ensemble of models described above, for each pair of branes $a$ and
$b$, we always obtain a pair of chiral matter in the bifundamental
representation of the two gauge group factors, coming from the intersection of
the two branes and the intersection of one brane with the orientifold mirror of
the other one.
One can define the two quantities
\begin{equation}
\label{eq:deltadef}
\Delta^\pm = |I_{ab}\pm I_{ab'}|,
\end{equation}
invariant under the exchange of branes and the orientifold action, that
describe the net amount of chiral matter for one particular brane.

We will use these two quantities as an example of how a correlation can arise
in the construction, and therefore in the amount of chiral matter that shows up
in the effective theory.
To show that this effect does not depend on the specific geometry, we compare
the analysis of the correlation pattern between intersecting brane models on
$T^6/(\mathbb{Z}_2\!\times\!\mathbb{Z}_2)$ and $T^6/\mathbb{Z}_6$ with a similar
analysis of Gepner model constructions~\cite{ci:timth}.
To see how far the actual distribution diverges from a generic match of
branes, we use a distribution with randomly generated pairings within the set
of branes of the model under consideration.

Although the distributions (Fig.~\ref{fig:corr}) are quite different
quantitatively, at a qualitative level the distribution of intersection numbers
is very similar for the intersecting brane models and the Gepner model
constructions. In particular the tendency towards either identical or rather
distant values for $\Delta^\pm$ is common in all three distributions.
This is quite remarkable, since one has to keep in mind that the Gepner models
are not only located at a different point in moduli space, but the ensemble
considered here consists of a sample of several thousand different Gepner models,
all corresponding to different backgrounds, of which only very few have even a
geometrical interpretation.

The fact that the distribution for $T^6/(\mathbb{Z}_2\!\times\!\mathbb{Z}_2)$
looks a bit blurred, can be traced back to the fact that the ensemble under
consideration has been cut off at high values of the complex structure
modulus, as explained above.
In the case of $T^6/\mathbb{Z}_6$ we are considering a random subset of
models that include also exceptional branes, which makes the ensemble
exponentially larger and at the same time reduces the number of possible values
for intersections~\cite{ci:z6}.

{\bf Acknowledgements} This work is supported by the Dutch Foundation for
Fundamental Research of Matter (FOM).

\end{document}